\documentclass[12pt]{article}
\usepackage{epsfig}
\usepackage{bbm,latexsym}
\usepackage{amssymb,amsmath}
\usepackage{slashed}
\newcommand{\bone}{\mathbbm{1}}
\newcommand{\dd}{\mathrm{d}}
\newtheorem{theorem}{Theorem}
\allowdisplaybreaks
\begin{document}
\title{
\normalsize \hfill UWThPh-2019-31 \\[10mm]
\LARGE Revisiting the quantum field theory of\\ neutrino oscillations in vacuum}

\author{
W.~Grimus\thanks{E-mail: walter.grimus@univie.ac.at}\;
\addtocounter{footnote}{1}
\\[5mm]
\small University of Vienna, Faculty of Physics \\
\small Boltzmanngasse 5, A--1090 Vienna, Austria
}

\date{January 10, 2020}

\maketitle

\begin{abstract}
We consider neutrino oscillations in vacuum in the framework of quantum 
field theory in which neutrino production and detection processes 
are part of a single Feynman diagram and the corresponding cross section 
is computed in the standard way, \textit{i.e.}\ with final states 
represented by plane waves. We use assumptions 
which are realized in actual experiments and concentrate 
on the detection process. Moreover, we also allow for a finite time interval 
of length $\tau$ during which the detector records neutrino events. 
In this context we are motivated by accelerator-neutrino oscillation 
experiments where data taking is synchronized in time with the 
proton spill time of the accelerator. 
Given the final momenta and the direction of neutrino 
propagation, we find that in the oscillation amplitude---for all 
practical purposes---the neutrino energy $Q$ is fixed, 
apart from an interval of order 
$2\pi\hbar/\tau$ around a mean energy $\bar Q$; this is an expression of 
energy non-conservation 
or the time-energy uncertainty relation
in the detection process due to $\tau < \infty$.
We derive in excellent approximation that in the amplitude 
the oscillation effect originates from massive neutrinos with the same 
energy $\bar Q$, \textit{i.e.}\ oscillations take place in space without any 
decoherece effect, while the remaining integration over $Q$ in the interval 
of order $2\pi\hbar/\tau$ around $\bar Q$ results in a 
time-correlation function expressing the time delay 
between neutrino production and detection. 
\end{abstract}

\newpage

\section{Introduction}
\label{introduction}
Neutrino oscillations have been proposed 
about 60 years ago~\cite{pontecorvo1,pontecorvo2} and are at present 
firmly established by a host of experiments~\cite{rpp}. However, the
theory of neutrino oscillations is still not a fully closed subject.
It is true that no thorough examination of the derivation of 
the standard neutrino oscillation 
probabilities~\cite{gribov,eliezer,fritzsch,bilenky1,bilenky2,bilenky3}
has ever cast doubts on their 
validity in realistic experiments---for reviews see 
\textit{e.g.}~\cite{bilenky-rev,petcov-rev,zralek,giunti-rev,beuthe,G2003}, but 
nevertheless some subtle issues, albeit of rather theoretical nature, remain 
to be clarified---for a recent paper see~\cite{akhmedov}. 
There are still two theoretical frameworks, which are kind of competing, 
namely the quantum-mechanical 
wave-packet 
approach~\cite{nussinov,kayser,giunti1991,lee,dolgov,akhmedov2009,naumov} 
and the quantum-field-theoretical 
approach~\cite{kobzarev,giunti1993,GS,giunti1997,GSM,ioannisian,kobach}. 
In~\cite{akhmedov2010} a 
comparison between the two approaches is carried out. 
In this paper we stick to the 
quantum-field-theory (QFT) framework because we think that it provides a 
more realistic interface between theory and experiment.  
This framework suggests to consider neutrino production and detection 
as a compound process~\cite{kobzarev,giunti1993,rich,campagne} 
included in a single Feynman diagram in which 
neutrinos propagating from the source to the detector 
are represented as inner lines.

The present paper can be considered as continuation of~\cite{G2003,GS,GSM}. 
(Here we do not take into account matter 
efffects~\cite{wolfenstein1,wolfenstein2,mikheyev,smirnov}.) 
We revisit the QFT derivation of neutrino oscillations for two reasons:
\begin{enumerate}
\renewcommand{\labelenumi}{\roman{enumi}.}
\item
In~\cite{G2003,GS,GSM} we have made the
unrealistic assumption that the detector particle is in an energy eigenstate.
In the present work we allow for an energy spread of the order of the thermal 
motion of the detector particles.\footnote{If different energies of the 
detector-particle state are not correlated, 
as advocated in~\cite{lipkin1,stodolsky,lipkin2}, 
then interference of states with different energies is not observable and  
the integration over the energy of the detector particle happens in the 
cross section, not in the amplitude. If this were the case, then,  
regarding the oscillation amplitude, this would effectively 
amount to the assumption of an energy eigenstate of the detector particle.
Since we are not sure about this issue, we consider the more general setting 
of admitting an energy spread of the detector particle in the oscillation 
amplitude.}
\item
One could be tempted to believe that neutrinos oscillate in time in 
accelerator experiments because, 
in order to reduce background, in such experiments
beam-induced neutrino events are correlated in time with the proton spill time 
of the accelerator~\cite{K2K,T2K,numi,nova}. 
Therefore, we simulate this situation by building in a finite time interval 
$\tau$ in the QFT oscillation amplitude in which the detector particle is ready 
for recording a neutrino. In practice, $\tau$ is between~1 and 
10\,$\mu$s.
\end{enumerate}

We employ a twofold strategy to tackle the QFT of neutrino oscillations.
On the one hand, 
we stress the necessity of using assumptions in our theoretical considerations
that are realized in actual experiments. 
On the other hand, we focus on the neutrino detection process since 
our analysis will show that this is the essential process 
concerning coherence in the neutrino oscillation amplitude, as 
demonstrated earlier in~\cite{weiss,kiers}.
At any rate, it will turn out that, using this strategy, 
neutrino oscillations follow naturally from QFT.
This includes the standard way of computing cross sections, 
\textit{i.e.}\ with plane waves in the final states.
Though for definiteness we will consider a model process for neutrino 
production and detection, our results will be of general validity.

We emphasize that in the present paper we are not concerned about 
decoherence effects in the oscillation probabilities originating from 
imperfect determination of the neutrino energy. Our subject is 
a possibly more fundamental decoherence effect in the oscillation 
\emph{amplitude} which could not be overcome even with perfect knowledge 
of the energies and momenta of all particles in the final state of the 
neutrino detection process.

In the paper we use natural units which are customary in QFT, 
but whenever we perform numerical estimates we insert Planck's constant 
$\hbar$ and the velocity of light $c$ in the appropriate places. 
In section~\ref{oscillations}, $\hbar$ and $c$ are employed throughout.

The plan of the paper is as follows. In section~\ref{hamiltonian}, in order 
to fix some notation, we write down the Hamiltonian in the $V-A$-theory and
in section~\ref{example} we introduce our model process and define its 
associated 4-momenta; moreover, we define the wave functions $\psi_S$ and 
$\psi_D$ of the neutrino source and detector particle, respectively, and 
discuss the functions in time which allow us to incorporate the finite 
lifetime of the source particle and the finite time interval for the 
neutrino flavour measurement in the detector. After having written 
down the full amplitude for the compound neutrino production--detection 
process in section~\ref{amplitude} and discussed the most suitable 
order of integrations in section~\ref{integrations}, we study in detail 
in section~\ref{detection} the relevant energies and 
the approximate energy conservation in the detection process. 
Section~\ref{asymptotic} is devoted to the asymptotic limit of large distance 
between source and detector and section~\ref{Qintegration} contains 
the discussion of the last remaining 
integration in the variable $Q$ which at this stage can be considered as the 
neutrino energy. We demonstrate in section~\ref{oscillations} 
that neutrino oscillations naturally follow 
from the QFT formalism. Finally, we summarize 
and present our conclusions in section~\ref{conclusions}.
Some mathematical tools, used in the body of the paper, are presented in the 
appendix.

\section{Hamiltonian}
\label{hamiltonian}
For simplicity, we assume that the average energy of the neutrino beam is 
well below the mass of the $W^\pm$ boson. Therefore, in neutrino production 
and detection we are allowed to use the $V - A$ Hamiltonian
\begin{equation}
H_{V-A} = \frac{G_F}{\sqrt{2}} \left( V^\mu - A^\mu \right)
\left( V_\mu - A_\mu \right)^\dagger
\end{equation}
with
\begin{equation}
V^\mu = \bar u \gamma^\mu  U_\mathrm{CKM} d + 
\bar\nu \gamma^\mu U_\mathrm{PMNS}^\dagger \ell,
\quad
A^\mu = \bar u \gamma^\mu \gamma_5 U_\mathrm{CKM} d + 
\bar\nu \gamma^\mu \gamma_5 U_\mathrm{PMNS}^\dagger \ell,
\end{equation}
where $U_\mathrm{CKM}$ is the Cabibbo--Kobayashi--Maskawa quark mixing 
matrix~\cite{kobayashi} 
and $U_\mathrm{PMNS}$ the Pontecorvo--Maki--Nakagawa--Sakata lepton mixing 
matrix~\cite{maki}. 
For a derivation of this Hamiltonian from the Standard Model
see for instance~\cite{horejsi}.
We can write the $V-A$ current as 
\begin{equation}
V^\mu - A^\mu = j^\mu_\mathrm{had} + j^\mu_\mathrm{lep} 
\end{equation}
with 
\begin{equation}
j^\mu_\mathrm{had} = 
\bar u \gamma^\mu \left( \bone - \gamma_5 \right) U_\mathrm{CKM} d, 
\quad \mbox{and} \quad
j^\mu_\mathrm{lep} = 
\bar\nu \gamma^\mu \left( \bone - \gamma_5 \right) U_\mathrm{PMNS}^\dagger \ell
\end{equation}
being the hadronic and the leptonic current, respectively. Furthermore, 
we use the notation 
\begin{equation}
U_\mathrm{PMNS} = \left( U_{\alpha j} \right)
\quad \mbox{with} \quad \alpha = e,\mu,\tau \; \mbox{and} \; j=1,2,3
\end{equation}
when we refer to elements of the PMNS matrix. The symbol $\nu$ denotes the 
vector of the three neutrino mass eigenfields $\nu_j$. In an analogous way 
we employ the symbols $u$, $d$ and $\ell$ for up-quarks, down-quarks and 
charged leptons, respectively. 

\section{Example and model process for neutrino production and detection}
\label{example}
We assume that $\beta$-decay of the nucleus $N = (Z,A)$ into the daughter 
nucleus $N' = (Z+1,A)$, \textit{i.e.}
\begin{equation}\label{S}
N(p) \to N'(p') + e^-(p'_e) + \bar\nu_e(q),
\end{equation}
constitutes the neutrino production process. The 4-momenta are indicated in 
parentheses. For  
neutrino detection, we assume the inverse $\beta$-decay
\begin{equation}\label{D}
\bar\nu_e(q) + p(k) \to n(k') + e^+(k'_e),
\end{equation}
where the neutron and the positron are measured in coincidence. 

We remind the reader that we consider 
neutrino source and detection processes as a single 
compound process. Therefore, the flavour neutrino $\bar\nu_e$ 
indicates that the production vertex is connected to the detection vertex
by neutrino propagators with (virtual) momentum $q$ and mass $m_j$. 

Though we have already fixed the notation of our 4-momenta, we have 
to take into account that we will frequently need energies and spatial 
momenta separately. 
The 4-momenta of $N$ and the proton are decomposed as 
\begin{equation}
p = \left( \begin{array}{c} E_S \\ \vec p \end{array} \right)
\quad \mbox{and} \quad
k = \left( \begin{array}{c} E_D \\ \vec k \end{array} \right),
\end{equation}
respectively,\footnote{Note that the meaning of the letter $p$ is 
ambiguous because it is also used to indicate the proton.}
where
\begin{equation}
E_S = \sqrt{m_N^2 + {\vec p}^{\,2}} 
\quad \mbox{and} \quad
E_D = \sqrt{m_p^2 + {\vec k}^{\,2}}. 
\end{equation}
Furthermore, 
\begin{equation}
p' = \left( \begin{array}{c} E'_S \\ {\vec p}^{\,\prime} \end{array} \right),
\quad
p'_e = \left( \begin{array}{c} E'_{eS} \\ {\vec p}_e^{\,\prime} \end{array} 
\right),
\quad
k' = \left( \begin{array}{c} E'_D \\ {\vec k}^{\,\prime} \end{array} \right),
\quad
k'_e = \left( \begin{array}{c} E'_{eD} \\ {\vec k}_e^{\,\prime} \end{array} 
\right)
\end{equation}
with
\begin{equation}
\begin{array}{ll}
E'_S = \sqrt{m_{N'}^2 + \left( {\vec p}^{\,\prime} \right)^2}, &
E'_{eS} = \sqrt{m_e^2 + \left( {\vec p}_e^{\,\prime} \right)^2}, \\[2mm]
E'_D = \sqrt{m_n^2 + \left( {\vec k}^{\,\prime} \right)^2}, &
E'_{eD} = \sqrt{m_e^2 + \left( {\vec k}_e^{\,\prime} \right)^2}.
\end{array}
\end{equation}
Note that all momenta associated with the neutrino source process have the 
letter $p$, while those associated with the detection process have $k$. 
All final momenta are primed, in contrast to the unprimed incoming ones.
Energies associated with the source process carry the subscript $S$, 
while those associated with the detection process have $D$. 

The state of the source particle $N$ is assumed to be smeared out in
momentum space by $\psi_S(\vec p\,)$ and that of the detector particle $p$ 
by $\psi_D(\vec k\,)$. Moreover, we assume that the functional forms of 
both $\psi_S(\vec p\,)$ and $\psi_D(\vec k\,)$ are such that they 
correspond to a localization of source and detector particle around 
$\vec x = \vec 0$ in coordinate space. 
The shifts to the space-time locations 
$(\vec x_S, t_S)$ and $(\vec x_D, t_D)$
of the source and detection process, respectively,  
will be performed in the compound amplitude $\mathcal{A}$ in the next section.

Note that the precise meaning of the localization of the 
\emph{unshifted} source and detector 
particles at $\vec x = \vec 0$ at times $t_S$ and $t_D$, respectively, is not 
important because, for reasonable functions 
$\psi_S(\vec p\,)$ and $\psi_D(\vec k\,)$ and reasonable widths of these 
functions---\textit{c.f.}\ section~\ref{Qintegration}, 
the widths of the respective Fourier transforms 
$\hat\psi_S(\vec x\,)$ and $\hat\psi_D(\vec x\,)$ will be microscopic, 
while the separation $L = \left| \vec x_D - \vec x_S \right|$ 
between the source and detector process is macroscopic.
Of course, only for non-relativistic particles 
the functions 
$\hat\psi_S(\vec x\,)$ and $\hat\psi_D(\vec x\,)$
can be conceived as quantum-mechanical probability amplitudes in 
coordinate space. For relativistic particles 
one can take an appropriate density 
(electric charge, baryon number,\,\ldots) to define localization. 
But these densities will in general still be concentrated in a microscopic 
region around the origin, just as 
$\left| \hat\psi_S(\vec x\,) \right|^2$ and 
$\left| \hat\psi_D(\vec x\,) \right|^2$.

Since $N$ decays we introduce the time dependence 
\begin{equation}\label{St}
\Theta(t - t_S) \exp \left( -\Gamma (t-t_S)/2\gamma \right)
\end{equation}
for the neutrino source, where  
$t_S$ denotes the time when $N$ begins to decay and neutrinos are thus 
produced, 
$\Theta$ is the Heaviside function,
$\Gamma$ the decay width and
the time-dilation factor $\gamma$ is given by 
\begin{equation}
\gamma = \frac{1}{\sqrt{1 - \beta_N^2}} = \frac{E_S}{m_N},
\end{equation}
where $\beta_N$ is the velocity of the source particle $N$ in units of $c$.

Concerning the detector, we assume that it measures in a time interval of 
length $\tau$ centered around the time $t_M$ ($t_M > t_S$). Thus we introduce 
the function\footnote{This has some similarity with the introduction of the 
``time-dependent propagator'' in~\cite{egorov}. However, in that paper the 
authors eventually set $\tau = 0$ and use plane waves also for the 
neutrino source and detector particles. Finally, their conclusions seem to 
differ considerably from ours.}
\begin{equation}\label{Dt}
\Theta_\tau(t - t_M) = \left\{
\begin{array}{ccc}
1 & \mbox{for} & |t - t_M| <  \frac{\tau}{2}, \\
0 & \mbox{for} & |t - t_M| >  \frac{\tau}{2}, 
\end{array} \right.
\end{equation}
to model the time-dependent data-taking in the detector.
We have to distinguish between the time $t_M$, the ``measurement time,'' and 
the time $t_D$ which is the time when the state of the 
detector particle is described by $\psi_D$. Since all 
\begin{equation}\label{equivalent}
t_D \in [t_M - \tau/2,\, t_M + \tau/2 ]
\end{equation}
are equivalent, we finally have to average over $t_D$ in the cross section of 
the compound process.

We stress that the functions of equations~(\ref{St}) and~(\ref{Dt})
break translation invariance in time. Therefore, energy conservation 
in the production and detection processes can only hold approximately.

\section{Amplitude}
\label{amplitude}
We introduce, for the relevant expectation values of the hadronic currents, 
the notation 
\begin{eqnarray}
\label{JS}
\frac{G_F}{\sqrt{2}} \langle N'(p') | 
j^\mu_\mathrm{had}(x) | N(p) \rangle 
& \equiv & J^\mu_{h,S}(p',p)\,e^{i(p'-p) \cdot x} ,
\\
\frac{G_F}{\sqrt{2}} \langle n(k') | 
\left( j^\mu_\mathrm{had}(x) \right)^\dagger | p(k) \rangle 
& \equiv & J^\mu_{h,D}(k',k)\,e^{i(k'-k) \cdot x} .
\label{JD}
\end{eqnarray}
In these expressions all particles are energy-momentum eigenstates.
The functions $\psi_S(\vec p\,)$ and $\psi_D(\vec k\,)$ will be taken 
into account below in the amplitude $\mathcal{A}$ of the compound process.

In order to formulate that the source particle is located at $\vec x_S$ at 
the time $t_S$ and that the detector particle is located at $\vec x_D$ at 
the time $t_D$, it is convenient to define the 4-vectors
\begin{equation}
x_S = \left( \begin{array}{c} t_S \\ \vec x_S \end{array} \right)
\quad \mbox{and} \quad 
x_D = \left( \begin{array}{c} t_D \\ \vec x_D \end{array} \right).
\end{equation}
In lowest order the compound amplitude $\mathcal{A}$ has two vertices,
one for the source process associated with the space-time variable $x_1$, 
and one for the detection process associated with $x_2$. 

Now we are in a position to put together the compound amplitude:
\begin{subequations}\label{A}
\begin{eqnarray}
\mathcal{A} &=& -i \sum_{j=1}^3 \int \dd^4 x_1 \int \dd^4 x_2 
\int \frac{\dd^4 q}{(2\pi)^4} e^{-iq \cdot (x_1 - x_2)}
\\ &&
\times 
\int \dd^3p\, \psi_S(\vec p\,) 
\exp \left[ -i p \cdot ( x_1 - x_S ) -
\Gamma (t_1-t_S)/(2\gamma) \right] \Theta(t_1 - t_S)
\\ &&
\times e^{i \left( p' + p'_e \right)\cdot x_1} J^\lambda_{h,S}(p',p) 
\\ &&
\times
\int \dd^3k\, \psi_D(\vec k\,) 
\exp \left[ - i k \cdot ( x_2 - x_D ) \right]
\Theta_\tau(t_2 - t_M)
\\ &&
\times e^{i \left( k' + k'_e \right)\cdot x_2} J^\rho_{h,D}(k',k) 
\\ &&
\times U_{ej} U^*_{ej}\,
\bar u_e(p'_e) \gamma_\lambda \left( \bone - \gamma_5 \right) 
\frac{\slashed{q} + m_j}{q^2 - m_j^2 + i\epsilon}
\gamma_\rho \left( \bone - \gamma_5 \right) v_e(k'_e).
\end{eqnarray}
\end{subequations}
In the last line, $u_e$ and $v_e$ are the 4-spinors of the electron and 
positron, respectively. The shift of the source and detector particles 
from $x = 0$ to $x_S$ and $x_D$, respectively, has been accomplished by 
the replacements
\begin{equation}
e^{-ip \cdot x_1} \to e^{-ip \cdot \left( x_1 - x_S \right)} 
\quad \mbox{and} \quad
e^{-ip \cdot x_2} \to e^{-ip \cdot \left( x_2 - x_D \right)},
\end{equation}
\textit{cf.}\ equations~(\ref{JS}) and~(\ref{JD}).
Therefore, $\psi_S(\vec p\,)$ and $\psi_D(\vec k\,)$ refer to wave functions 
in momentum space at times $t_S$ and $t_D$, respectively.

\section{Integrations}
\label{integrations}
To proceed further in a transparent way, 
the order of the integrations is crucial:
\begin{enumerate}
\item 
$\int \dd^3 x_1$ and $\int \dd^3 x_2$,
\item
$\int \dd^3p$ and $\int \dd^3k$,
\item
$\int \dd t_1$ and $\int \dd t_2$,
\item
$\int \dd^3 q$ in the asymptotic limit 
$L \equiv \left| \vec x_D - \vec x_S \right| \to \infty$,
\item
$\int \dd q^0$ in an approximation to be discussed.
\end{enumerate}
While the first three integrations are elementary, the last two can only 
be performed in suitable approximations and will be treated in separate 
sections. 

The first integration leads to $\delta$-functions:
\begin{equation}
(2\pi)^6 \,\delta(\vec p + \vec q - {\vec p}^{\,\prime} - {\vec p}_e^{\,\prime})
\, \delta(\vec k - \vec q - {\vec k}^\prime - {\vec k}_e^\prime).
\end{equation}
Therefore, the second integration amounts to the replacements
\begin{equation}\label{second int}
\vec p  \to \tilde{\vec p} \equiv 
{\vec p}^{\,\prime} + {\vec p}_e^{\,\prime} - \vec q,
\quad
\vec k \to \tilde{\vec k} \equiv 
{\vec k}^\prime + {\vec k}_e^\prime + \vec q
\end{equation}
everywhere in $\mathcal{A}$. In order to keep track of these replacements
in the energies as well, we define 
\begin{equation}
\tilde E_S = \sqrt{ m_N^2 + \left( \tilde{\vec p} \right)^2},
\quad
\tilde E_D = \sqrt{ m_p^2 + \left( \tilde{\vec k} \right)^2},
\quad
\tilde \gamma = \tilde E_S/m_N,
\end{equation}
and
\begin{equation}
\tilde p = \left( \begin{array}{c} \tilde E_S \\ \tilde{\vec p}
\end{array} \right),
\quad
\tilde k = \left( \begin{array}{c} \tilde E_D \\ \tilde{\vec k}
\end{array} \right).
\end{equation}
The time integrations can be performed explicitly as well. In summary, 
after three integration steps the amplitude has the form 
\begin{subequations}\label{A3}
\begin{eqnarray}
\mathcal{A} &=& -i (2\pi)^3 \times e^{i(p'+p'_e) \cdot x_S} \times 
e^{i(k'+k'_e) \cdot x_D} \times \sum_{j=1}^3 \int \dd^4 q\, e^{iq \cdot (x_D - x_S)}
\\ &&
\times \psi_S(\tilde{\vec p}\,)  J^\lambda_{h,S}(p',\tilde p) 
\times \psi_D(\tilde{\vec k}\,)  J^\rho_{h,D}(k',\tilde k) 
\\ &&
\times \frac{1}{i \left( q^0 + \Delta \tilde E_S \right) + 
\Gamma/(2\tilde \gamma)} 
\times \frac{\sin\left( \frac{\tau}{2}( q^0 - \Delta \tilde E_D ) \right)}%
{\pi( q^0 - \Delta \tilde E_D)}
\,
e^{i(q^0 - \Delta \tilde E_D)(t_M - t_D)}
\label{tau} \\ &&
\times U_{ej} U^*_{ej}\,
\bar u_e(p'_e) \gamma_\lambda \left( \bone - \gamma_5 \right) 
\frac{\slashed{q} + m_j}{q^2 - m_j^2 + i\epsilon}
\gamma_\rho \left( \bone - \gamma_5 \right) v_e(k'_e).
\end{eqnarray}
\end{subequations}
For simplicity of notation we have introduced the energy differences
\begin{equation}\label{Delta E}
\Delta \tilde E_S = \tilde E_S - E'_S - E'_{eS}
\quad \mbox{and} \quad
\Delta \tilde E_D = \tilde E_D - E'_D - E'_{eD}.
\end{equation}

\section{The detection process}
\label{detection}
Before we proceed further with the final integrations, it is 
useful consider the detection process in more detail. This will help us
to understand the relevant approximations.

\paragraph{Approximate energy conservation in neutrino detection:}
The amplitude $\mathcal{A}$ depends on the measurement interval 
$\tau$ as can be gathered from equation~(\ref{tau}).
In the limit of infinitely long measurement time we obtain exact 
energy conservation in the measurement reaction, \textit{i.e.}\ 
\begin{equation}\label{tau-infty}
\lim_{\tau \to \infty}
\frac{\sin\left( \frac{\tau}{2}( q^0 - \Delta \tilde E_D ) \right)}%
{\pi( q^0 - \Delta \tilde E_D)} = 
\delta \left( q^0 - \Delta \tilde E_D \right).
\end{equation}
The symbol ``$\delta$'' in this equation denotes 
the one-dimensional delta function.

In accelerator experiments, $\tau$ is not infinitely long but of 
the order of microseconds and 
energy will only approximately be conserved in the detection 
reaction.\footnote{It appears strange that selecting detector events 
from a specific time interval introduces a tiny energy non-conservation,
however, this can be conceived as an expression of the 
time-energy uncertainty relation.}
Inserting $\hbar$ into the sine function, 
the approximate range of $q^0$ is determined by 
the inequality 
\begin{equation}
\frac{\tau}{2\hbar}
\left| q^0 - \Delta \tilde E_D \right| \lesssim
\pi,
\end{equation}
where the right-hand side indicates the first zero of the 
sine function.\footnote{More realistically, we should take some multiple of 
$\pi$. We will do this later in section~\ref{oscillations}, where we discuss 
an approximate integral over $Q$.} In all numerical considerations involving 
$\tau$ we will set $\tau = 10^{-6}$\,s, which is in the right ballpark used in 
experiments~\cite{K2K,T2K,numi,nova}.
Then, with $\hbar \simeq 6.582 \times 10^{-22}$\,MeV\,s 
we obtain 
\begin{equation}\label{de}
\left| q^0 - \Delta \tilde E_D \right| \lesssim
\frac{2\pi \hbar}{\tau} \simeq 4 \times 10^{-9}\,\mbox{eV}.
\end{equation}
We conclude that---with a realistic value of $\tau$---approximate 
energy conservation in the detection reaction 
is much better than any energy measurement accuracy in particle scattering.

\paragraph{Sign and order of magnitude of $q^0$:}
Let us recall realistic neutrino-detection 
reactions. One possibility is a general inverse $\beta$-decay 
with a \emph{stable} detector particle, where the 
charged lepton in the final state uniquely indicates the neutrino flavour.
The other one is elastic neutrino--electron scattering, without unique 
neutrino-flavour determination. Moreover, in practice, 
the detector particle in these reactions is at rest.
Of course, it is at rest only for all practical purposes because in a fluid 
it undergoes thermal motion and in a solid body it oscillates around its 
equilibrium position. 

In the case of a general inverse $\beta$-decay, 
stability of the detector particle means, in particular, 
that it cannot have $\beta$-decay. Therefore, 
neglecting the small neutrino masses,
the mass of the detector particle is smaller than the sum 
over the masses in the final state of the detection process, \textit{i.e.}\ 
there is an energy threshold to overcome. Usually it is necessary to observe 
the track of the final charged lepton in the detector, then, in addition, 
the kinetic energy of the charged lepton must be 
above a certain minimum.

In our model detection reaction 
$\bar\nu_e + p \to e^+ + n$ this threshold is given by the well-known value 
$m_n + m_e - m_p \simeq 1.8$\,MeV. Since here the positron and the neutron 
are measured in coincidence, no minimum kinetic energy of the positron is 
required. 

In elastic neutrino-electron scattering as the detection reaction 
the kinetic energy of the electron in the final state  
must be at least several 100\,keV, 
much larger than the energy of thermal motion, 
in order to have reasonable background rejection. 

Putting these rather trivial points together, we conclude that 
$\Delta \tilde E_D$ of equation~(\ref{Delta E}) or any analogon thereof 
corresponding to other detection reactions must be negative and
and its absolute value of the order of several 100\,keV or more. 
For definiteness we set
\begin{equation}
-\Delta \tilde E_D \gtrsim 0.5\,\mbox{MeV},
\end{equation}
which we will use in the following for numerical estimates.
Consequently, because of the approximate energy conservation expressed 
by equation~(\ref{de}), we find that only values of $q^0$ which fulfill 
\begin{equation}\label{q0}
q^0 < 0 \quad \mbox{and} \quad -q^0 \gtrsim 0.5\,\mbox{MeV}
\end{equation}
can contribute 
significantly to $\mathcal{A}$.\footnote{In general, 
for antineutrino detection one finds $q^0 < 0$ while $q^0 > 0$ holds
for neutrino detection.}
This will be relevant in the next section.

\section{Integration over $\dd^3 q$ and asymptotic limit $L \to \infty$}
\label{asymptotic}
The integration $\int \dd^3 q$ cannot be performed exactly, but anyway we 
need the result only in the asymptotic limit 
$L = |\vec x_D - \vec x_S | \to \infty$. 
For this purpose we use the theorem 
proven in~\cite{GS} that is reproduced 
in the present paper in appendix~A as theorem~\ref{limL}.

In order to simplify our notation, 
we define, in view of equation~(\ref{q0}), the positive quantity 
\begin{equation}
Q = -q^0.
\end{equation}
Then, 
theorem~\ref{limL} 
tells us that, in the asymptotic limit $L \to \infty$, 
the dominant contributions to $\mathcal{A}$, equation~(\ref{A3}),
require $Q^2 > m_j^2$ $\forall j$ and drop as $1/L$,
while subdominant terms diminish at least as $1/L^{3/2}$.
In order to conveniently formulate the dominant terms as given 
by theorem~\ref{limL}, 
we define 
\begin{equation}\label{qj}
\vec q_j = \sqrt{Q^2 - m_j^2} \, \vec \ell
\quad \mbox{and} \quad 
q_j = \left( \begin{array}{c} Q \\ \vec q_j
\end{array} \right)
\end{equation}
with 
\begin{equation}\label{vec ell}
\vec L = \vec x_D -  \vec x_S
\quad \mbox{and} \quad
\vec \ell = \frac{\vec L}{L}.
\end{equation}
Note that 
the 4-momentum vectors $q_j$ are on-mass-shell:
\begin{equation}
q_j^2 = m_j^2.
\end{equation}
Denoting the amplitude in the asymptotic limit by $\mathcal{A}_\infty$ and 
invoking once more theorem~\ref{limL}, 
we obtain the result
\begin{subequations}\label{Ainfty}
\begin{eqnarray}
\mathcal{A}_\infty &=& -i (2\pi)^3 \times e^{i(p'+p'_e) \cdot x_S} \times 
e^{i(k'+k'_e) \cdot x_D} \times \left( -\frac{2\pi^2}{L} \right) 
\nonumber \\ && \times
\sum_{j=1}^3 \int \dd Q\, \Theta\left( Q^2 - m_j^2 \right) 
e^{-iQ(t_D - t_S)+ i\sqrt{Q^2 - m_j^2} L}
\label{exp} \\ && 
\times \left. \psi_S(\tilde{\vec p}\,)  J^\lambda_{h,S}(p',\tilde p) 
\right|_{q = -q_j}
\times \left. \psi_D(\tilde{\vec k}\,)  J^\rho_{h,D}(k',\tilde k) 
\right|_{q = -q_j}
\label{psiJ} \\ &&
\left. \times \frac{i}{Q - \Delta \tilde E_S + 
i\Gamma/(2\tilde \gamma)} \right|_{q = -q_j} 
\times \left. \frac{\sin\left( \frac{\tau}{2}( Q + \Delta \tilde E_D ) \right)}%
{\pi( Q + \Delta \tilde E_D)} 
\, e^{-i \left( Q + \Delta \tilde E_D \right) \left( t_M - t_D \right)}
\right|_{q = -q_j} 
\label{time} \\ &&
\times U_{ej} U^*_{ej}\,
\bar u_e(p'_e) \gamma_\lambda \left( \bone - \gamma_5 \right) 
\left( -\slashed{q}_j + m_j \right)
\gamma_\rho \left( \bone - \gamma_5 \right) v_e(k'_e).
\end{eqnarray}
\end{subequations}

In this asymptotic limit the intermediate virtual neutrinos become
on-mass-shell, which allows us to use the decomposition
\begin{equation}\label{decomp}
{\slashed{q}}_{\!j} - m_j = \sum_s v(q_j,s) \bar v(q_j,s),
\end{equation}
where $s$ denotes the two independent spin states of the antineutrino.
As a consequence, equation~(\ref{Ainfty}) can be written as 
\begin{subequations}\label{Ainf}
\begin{eqnarray}
\mathcal{A}_\infty &=& \frac{1}{L}\, 
\sum_{j=1}^3  U_{ej} U^*_{ej}\, \int \dd Q\, \Theta\left( Q^2 - m_j^2 \right) 
e^{-iQ(t_D - t_S)+ i\sqrt{Q^2 - m_j^2} L}
\label{Ainf-exp} \\ && 
\times
\mathcal{A}_S(p',p'_e,q_j)\,
\mathcal{A}_D(k',k'_e,q_j) 
\\ && \times 
\left. \frac{\sin\left( \frac{\tau}{2}( Q + \Delta \tilde E_D ) \right)}%
{\pi( Q + \Delta \tilde E_D)} 
\, e^{-i \left( Q + \Delta \tilde E_D \right) \left( t_M - t_D \right)}
\right|_{q = -q_j}.
\label{Ainf-amp}
\end{eqnarray}
\end{subequations}
The amplitudes $\mathcal{A}_S(p',p'_e,q_j)$ and $\mathcal{A}_D(k',k'_e,q_j)$
refer to the neutrino source and detection reaction, 
respectively. These amplitudes are unique up to factors which play no role 
in the following discussion:
\begin{subequations}
\begin{eqnarray}
\mathcal{A}_S(p',p'_e,q_j) &\propto&
\left. \psi_S(\tilde{\vec p}\,)  J^\lambda_{h,S}(p',\tilde p) 
\right|_{q = -q_j} \times 
\left. \frac{i}{Q - \Delta \tilde E_S + 
i\Gamma/(2\tilde \gamma)} \right|_{q = -q_j} 
\nonumber \\ && \times
\bar u_e(p'_e) \gamma_\lambda \left( \bone - \gamma_5 \right) v(q_j,+),
\label{AS} 
\\
\mathcal{A}_D(k',k'_e,q_j) &\propto&
\left. \psi_D(\tilde{\vec k}\,)  J^\rho_{h,D}(k',\tilde k) 
\right|_{q = -q_j} \times 
\bar v(q_j,+) \gamma_\rho \left( \bone - \gamma_5 \right) v_e(k'_e).
\label{AD}
\end{eqnarray}
\end{subequations}
In the 4-spinor $v(q_j,+)$ we have indicated the positive helicity of the 
antineutrino.

In view of the discussion in the previous section, in equation~(\ref{Ainf}) 
the integration over negative $Q$ is completely negligible. 
In addition, only values of $Q$ with $Q \gg m_j$ are relevant.
Moreover, equation~(\ref{qj}) suggests to consider 
$Q$ simply as the neutrino energy.

Finally, it is expedient to discuss the condition for the 
applicability of the asymptotic limit 
given by 
theorem~\ref{limL}. 
Inserting $\hbar c$, this condition reads
\begin{equation}
\frac{QL}{\hbar c} \gg 2\pi.
\end{equation}
Taking as an example $Q = 0.5$\,MeV and $L = 300$\,km, we find
\begin{equation}
\frac{QL}{\hbar c} \simeq 0.8 \times 10^{18}.
\end{equation}
Thus for all realistic situations the asymptotic limit is justified.

\section{Integration over $\dd Q$}
\label{Qintegration}
\paragraph{Estimating the widths of $\psi_D$ and $\psi_S$:}
These wave functions describe the momentum distributions of 
the neutrino-dector and source particles, respectively. 
Since in our discussion we put more weight on the neutrino-detection 
reaction than on the neutrino-source reaction, we consider 
first the detector particle, which in our model reaction 
is the proton. As stated in the introduction, we assume that its
non-relativistic thermal motion, \textit{i.e.}\ the mean value 
$\langle E_{\mathrm{kin},D} \rangle$ of its kinetic energy,
gives the correct ballpark estimate 
of its energy spread. Therefore, the width of $\psi_D$ denoted by $\sigma_D$
is estimated in the following way:
\begin{equation}
\langle E_{\mathrm{kin},D} \rangle = \frac{3}{2} kT 
\quad  \Rightarrow \quad
\sigma_D \sim \sqrt{2m_p \langle E_{\mathrm{kin},D} \rangle} = \sqrt{3m_pkT}.
\end{equation}
Inserting $kT \simeq (1/40)$\,eV for room temperature, we obtain 
the order-of-magnitude estimate
\begin{equation}\label{sigmaD}
\sigma_D \sim 10\,\mbox{keV}.
\end{equation}

Since in our model reaction the neutrino comes from some $\beta$-decay of 
a nucleus at rest, $\sigma_S$, the width of $\psi_S$, could be an order of 
magnitude larger. If the neutrino source, for instance charged pions, 
decays in flight, we would assume that $\sigma_S$ is much larger.
Whatever the real value of $\sigma_S$ is, in the following only 
\begin{equation}\label{sigmaSD}
\sigma_S \gtrsim s_D
\end{equation}
will be relevant.

\paragraph{The limit $\tau \to \infty$:}
As discussed in section~\ref{detection}, in this limit we have exact 
energy conservation in the detection process:
\begin{equation}\label{Econs}
Q + \left. \Delta \tilde E_D \right|_{q = -q_j} = 0
\quad \mbox{with} \quad
\left. \Delta \tilde E_D \right|_{q = -q_j} =  \left. 
\tilde E_D \right|_{q = -q_j} - E'_D - E'_{eD}.
\end{equation}
This equation depends on the neutrino mass $m_j$. Thus, for every $j=1,2,3$
there will be a separate solution $Q_j$ of equation~(\ref{Econs}). 
Knowing that $Q \gtrsim 0.5$\,MeV, we are allowed to perform 
an expansion in $m_j^2$:
\begin{subequations}
\begin{eqnarray}
\left. \tilde E_D \right|_{q = -q_j} &=& \left[ m_p^2 + \left( 
\vec k' + \vec k'_e -\sqrt{Q^2 - m_j^2}\, \vec\ell\, \right)^2
\,\right]^{1/2}
\label{EDtilde} \\ &\simeq&
\left[ m_p^2 + \left( \vec k' + \vec k'_e - Q \vec\ell \,\right)^2 \right]^{1/2} 
+ \frac{m_j^2}{2 m_p Q}\,\vec \ell \cdot 
\left( \vec k' + \vec k'_e - Q \vec \ell \,\right).
\label{m-corr}
\end{eqnarray}
\end{subequations}
Equation~(\ref{m-corr}) immediately leads to the approximate solution 
$Q_j$ of equation~(\ref{Econs}) given by 
\begin{equation}\label{Q_j}
Q_j = \bar Q - \frac{m_j^2}{2 m_p \bar Q}\,\vec \ell \cdot \vec {\bar k}
\quad \mbox{with} \quad 
\vec {\bar k} = \vec k' + \vec k'_e - \bar Q \vec \ell,
\end{equation}
where $\bar Q$ is the solution of
\begin{equation}\label{Qbar1}
\bar Q + \left[ m_p^2 + \left( \vec {\bar k}\, \right)^2
\right]^{1/2} - E'_D - E'_{eD} = 0.
\end{equation}

Note that, in the approximation $m_j = 0$, 
$\vec {\bar k}$ is the momentum of the detector particle picked out by 
the momentum configuration at hand, defined by $\vec\ell$, 
$\bar Q$, $\vec k'$ and $\vec k'_e$---see 
equation~(\ref{second int}). Therefore, 
\begin{equation}\label{kD}
\left| \vec {\bar k}\, \right| \lesssim \sigma_D.
\end{equation}

Let us estimate the order of magnitude of the term proportional to $m_j^2$ in 
equation~(\ref{Q_j}). Taking into account equation~(\ref{kD}), 
$Q \gtrsim 0.5$\,MeV and $m_j \lesssim 0.1$\,eV~\cite{rpp}, this term 
is of order $10^{-13}$\,eV or smaller, for the detector particle being a 
proton.\footnote{If the detector particle is an electron, 
this upper limit would be larger by a factor of 2000, but the mass correction 
would still be negligible.}
Thus, for all practical purposes, we can set $m_j = 0$ in 
equation~(\ref{Q_j}) and use $Q_j = \bar Q$ for $j=1,2,3$ 
in the context of neutrino oscillations. 

In effect, with $m_j = 0$ we infer from equation~(\ref{Econs}) that, 
with fixed energies of the final particles in the detection process, the 
detector particle has a definite energy. 
This is exactly what has been 
\emph{assumed} in~\cite{GS,GSM}, while here we have \emph{demonstrated}
that this is correct with excellent accuracy. 

As a side remark and anticipating neutrino oscillations, 
using the $m_j$-dependent $Q_j$ of equation~(\ref{Q_j}) 
in $\mathcal{A}_\infty$ of equation~(\ref{Ainf}), 
we would introduce tiny, in practice irrelevant 
oscillations in time, \textit{i.e.} in $t_D - t_S$,  
in addition to those in space, \textit{i.e.}\ in $L$.

Finally, we discuss the delta function of equation~(\ref{tau-infty}), 
referring to energy conservation in the detection process, in the 
approximation $m_j = 0$. For this purpose we define the 4-momentum
\begin{equation}\label{qbar}
\check q(Q) = Q \left( \begin{array}{c} 1 \\ \vec \ell 
\end{array} \right),
\end{equation}
which is obtained from $q_j$ by setting $m_j = 0$. Then, linear expansion 
in $Q - \bar Q$ of the energy conservation relation  gives
\begin{equation}\label{linear}
Q + \left. \Delta \tilde E_D \right|_{q = -\check q(Q)} = 
(1 - \beta) (Q - \bar Q)
\end{equation}
with
\begin{equation}\label{beta}
\beta = 
-\left. \frac{\left. \partial \tilde E_D \right|_{q = -\check q(Q)}}%
{\partial Q} \right|_{Q = \bar Q} =
\frac{\vec {\bar k} \cdot \vec \ell}{\tilde {\bar E}_D}
\quad \mbox{and} \quad 
\tilde {\bar E}_D = \left. \tilde E_D \right|_{q = -\check q(\bar Q)}.
\end{equation}
Therefore, in the approximation $m_j = 0$, the delta function of 
equation~(\ref{tau-infty}) reads
\begin{equation}
\delta \left( Q + \left. \Delta \tilde E_D \right|_{q = -\check q(Q)} \right) = 
\frac{1}{1-\beta}\,\delta \left( Q - \bar Q \right).
\end{equation}
Evidently, for a given momentum configuration, 
$\beta$ is the projection of the velocity of the detector 
particle unto the direction $\vec\ell$ in units of $c$.
Since the detector particle is at rest apart from 
thermal motion, we have $|\beta| \ll 1$. In the case of the proton, with 
equations~(\ref{sigmaD}) and~(\ref{kD}) we find 
$|\beta| \lesssim 10^{-5}$.

\paragraph{Finite $\tau$ and the relevant range of $Q$:}
In this case the neutrino energy is not fixed. However, we will have 
approximate energy conservation in the detection process---\textit{cf.}\ 
section~\ref{detection}---expressed as
\begin{equation}\label{relevant range}
\left| Q + \left. \Delta \tilde E_D \right|_{q = -q_j} \right|
\lesssim \frac{2\pi\hbar}{\tau}.
\end{equation}
Note that the neutrino energy $Q$ is not only restricted 
by the $\tau$-dependent factor in equation~(\ref{Ainf-amp}) but also 
by its occurence in $\psi_S$ and $\psi_D$ 
in $\mathcal{A}_S$ of equation~(\ref{AS}) and
in $\mathcal{A}_D$ of equation~(\ref{AD}), respectively.
(Its role in the exponents of equation~(\ref{Ainf}) will be discussed 
in the next section, when we treat neutrino oscillations.) 
However, numerically the width $\sigma_D$ is much larger than 
$(2\pi \hbar)/\tau$---compare equation~(\ref{de}) with 
equation~(\ref{sigmaD}). Therefore, we have 
\begin{equation}
\sigma_D \gg \frac{2\pi \hbar}{\tau}.
\end{equation}
Finally, taking into account equation~(\ref{sigmaSD}),
we find that indeed equation~(\ref{relevant range})
gives the strongest restriction on $Q$.

In the further discussion of equation~(\ref{relevant range}) we 
set $m_j = 0$.
According to equation~(\ref{de}), 
the relevant integration interval of the variable $Q$ extends, 
for all practical purposes, only over a 
tiny interval of a length $\Delta Q$ not much larger than $10^{-8}$\,eV. 
As stressed in section~\ref{detection}, the reason is that 
the $\tau$-dependent factor in equation~(\ref{Ainf-amp}) 
is almost a $\delta$-function. 
Therefore, we decompose $Q$ as
\begin{equation}\label{Qbar0}
Q = \bar Q + \delta Q,
\end{equation}
where $\bar Q$ is the mean value of the relevant range of $Q$. 
Thus we have 
\begin{equation}
-\frac{1}{2} \Delta Q \lesssim \delta Q \lesssim \frac{1}{2} \Delta Q
\quad \mbox{and} \quad 
\Delta Q \sim \frac{2\pi \hbar}{\tau}.
\end{equation}
The mean energy $\bar Q$ is determined by equation~(\ref{Qbar1}).

\paragraph{Computation of $\bar Q$:}
Let us now compute $\bar Q$ approximately by 
expanding $\bar Q$ in powers of $1/m_p$ as
\begin{equation}\label{Qbar}
\bar Q = Q_0 + \Delta_1 + \Delta_2 + \Delta_3 + \ldots
\quad \mbox{with} \quad Q_0 = E'_D + E'_{eD} - m_p.
\end{equation}
This expansion originates in an expansion of the square root in 
equation~(\ref{Qbar1}), which makes sense because of equation~(\ref{kD}) and 
$\sigma_D^2/m_p^2 \sim 10^{-10}$.
The lowest order $Q_0$ corresponds to a proton at rest without thermal motion.
Defining
\begin{equation}
\vec q_0 = \vec k' + \vec k'_e -  Q_0 \vec\ell,
\end{equation}
we find
\begin{equation}
\Delta_1 = -\frac{{\vec q_0}^{\,2}}{2m_p}, \quad
\Delta_2 = -\frac{{\vec q_0}^{\,2}}{2m_p^2}\,\vec\ell \cdot \vec q_0, \quad
\Delta_3 = -\frac{{\vec q_0}^{\,2}}{2m_p^3}
\left( \vec\ell \cdot \vec q_0 \right)^2.
\end{equation}
The order of $\Delta_3$ is given by $\sigma_D^4/m_p^3 \sim 10^{-11}$\,eV.

\paragraph{An approximation for $\mathcal{A}_\infty$:}
The length $\Delta Q$ of the integration interval for $Q$ is very small.
So it is quite conceivable that, when $Q$ is in this interval,  
the $Q$-dependence of the amplitudes 
$\mathcal{A}_S$ and $\mathcal{A}_D$ will be totally negligible because 
$\Delta Q$ is much smaller than all relevant energies in these amplitudes. 
This statement requires, in particular, that the relevant scales on which 
the wave functions $\psi_S$ and $\psi_D$ vary, are much larger than $\Delta Q$. 
But this is a natural assumption due to 
$\sigma_S \gtrsim \sigma_D \gg \Delta Q$. Likewise, 
we can also neglect the neutrino masses in these amplitudes. Therefore, 
we can make the replacements 
\begin{equation}
q_j \to \check q(\bar Q) = \bar Q \left( \begin{array}{c} 1 \\ \vec \ell 
\end{array} \right)
\end{equation}
in $\mathcal{A}_S$ and $\mathcal{A}_D$. 
Denoting the resulting $Q$-independent amplitudes by 
$\bar{\mathcal{A}}_S$ and $\bar{\mathcal{A}}_D$, we extract them from the 
integral in equation~(\ref{Ainf}). Finally, switching from the integration 
variable $Q$ to $\delta Q$---see equation~(\ref{Qbar0}),
we obtain the approximation 
\begin{eqnarray}
\lefteqn{
\mathcal{A}_\infty \simeq \frac{1}{L}\, \bar{\mathcal{A}}_S \bar{\mathcal{A}}_D
\sum_{j=1}^3 U_{ej} U_{ej}^* \int_{-\Delta Q/2}^{\Delta Q/2} 
\dd \delta Q} \nonumber \\
&& \exp \left[ 
i \left( -\left (\bar Q +\delta Q \right)(t_D - t_S) + 
\sqrt{ \left( \bar Q +\delta Q \right)^2 - m_j^2}\, L \right) \right]
\nonumber \\ &&
\times 
\left. \frac{\sin\left( \frac{\tau}{2}( Q + \Delta \tilde E_D ) \right)}%
{\pi( Q + \Delta \tilde E_D)} 
\, e^{-i \left( Q + \Delta \tilde E_D \right) \left( t_M - t_D \right)}
\right|_{q = -\check q(\bar Q + \delta Q)}.
\label{A-approx}
\end{eqnarray}
The discussion of neutrino oscillations in the next section will based on
this form of $\mathcal{A}_\infty$.

\section{Neutrino oscillations}
\label{oscillations}
In this section, for the sake of clarity, we do not use natural units but 
insert $\hbar$ and $c$ in the appropriate places.
\paragraph{The oscillation phase:}
Denoting for each $j=1,2,3$ the collection of 
phases appearing in the exponents of equation~(\ref{A-approx}) 
by $\mathcal{E}_j$, we have 
\begin{subequations}
\begin{eqnarray}
\mathcal{E}_j &=& \frac{1}{\hbar} \left[
-\left (\bar Q +\delta Q \right)(t_D - t_S) + 
\sqrt{ \left( \bar Q +\delta Q \right)^2/c^2 - (m_j c)^2}\, L \right.
\label{Ea} \\ && -
\left. \left. \left( Q + \Delta \tilde E_D \right) \left( t_M - t_D \right) 
\right|_{q = -\check q(\bar Q + \delta Q)}\, \right].
\label{Eb}
\end{eqnarray}
\end{subequations}
Clearly, neutrino masses must not be neglected in these phases 
because there the effect of $m_j^2$ is amplified by 
the huge macroscopic length $L$, which is the cause of observable 
neutrino oscillations. 
However, in view of $m_j \ll \bar Q$, $|\delta Q| \ll \bar Q$ 
and $|\delta Q| \ll m_p$, we perform 
an expansion in $1/\bar Q$ in equation~(\ref{Ea}) until $(1/\bar Q)^3$ 
and in equation~(\ref{Eb}) until $(\delta Q)^2$:
\begin{subequations}
\label{phase-exp}
\begin{eqnarray}
\mathcal{E}_j &\simeq& 
-\frac{\bar Q}{\hbar} \left(t_D - t_S - \frac{L}{c} 
\right) 
-\frac{(m_j c^2)^2 L}{2 \hbar c \bar Q} 
-\frac{(m_j c^2)^4 L}{8 \hbar c {\bar Q}^3} 
\label{phase1} \\ &&
-\frac{\delta Q}{\hbar} \left[ t_D - t_S - \frac{L}{c} 
- \frac{(m_j c^2)^2 L}{2 c {\bar Q}^2}
\left( 1 - \frac{\delta Q}{\bar Q} \right)
\right]
\label{phase2} \\ &&
-\frac{\delta Q}{\hbar} \left( t_M - t_D \right) \left[
1 - \beta + \frac{\delta Q}{2 \tilde{\bar E}_D} \left( 1 - \beta^2 \right)
\right].
\label{phase3}
\end{eqnarray}
\end{subequations}
Note that, in an expansion of $\mathcal{E}_j$ in $1/\bar Q$, terms 
$\frac{\bar Q L}{\hbar c} %
\left( \frac{\delta Q}{\raisebox{-1pt}{$\scriptstyle\bar Q$}} \right)^n$ with 
$n = 2,3,4,\dots$, \textit{i.e.}\ without powers of $m_j^2$, cannot occur. 
This can be inferred from setting $m_j = 0$ in 
the square root of $\mathcal{E}_j$. The term linear in $\delta Q$ 
in equation~(\ref{phase3}) has already been displayed in 
equation~(\ref{linear}), whereas the definitions of $\beta$ and 
$\tilde{\bar E}_D$ can be found in equation~(\ref{beta}). 
The expansion in equation~(\ref{phase-exp}) contains the dominant terms, 
which we will take into account in the following, plus corrections to 
the dominant terms, which will be neglected but allow an error estimation.

Obviously, the first term on the right-hand side of equation~(\ref{phase1}) 
drops out in the cross section and thus has no physical effect. 
The second term in equation~(\ref{phase1}) gives rise to neutrino oscillations 
with oscillation lengths proporional to inverse 
mass-squared differences $m_i^2 - m_j^2$, while the third term is 
a correction to the second one but irrelevant in practice. 

\paragraph{Approximations:}
To proceed with the amplitude of equation~(\ref{A-approx}), 
we make the following approximations:
\begin{enumerate}
\renewcommand{\labelenumi}{\roman{enumi}.}
\item
We replace the phase by its expanded form, equation~(\ref{phase-exp}).
\item
As announced above, 
we simplify $\mathcal{E}_j$ further by dropping the third term in 
equation~(\ref{phase1}) and by neglecting the terms quadratic in 
$\delta Q$ in equations~(\ref{phase2}) and~(\ref{phase3}).
\item
Finally, in the integral of equation~(\ref{A-approx}) we take the limit 
$\Delta Q \to \infty$.
\end{enumerate}
In summary, with these approximations, equation~(\ref{A-approx}) reads
\begin{subequations}
\label{A-int}
\begin{eqnarray}
\lefteqn{\mathcal{A}_\infty \simeq 
\frac{1}{L}\, \bar{\mathcal{A}}_S \bar{\mathcal{A}}_D
} \nonumber \\ && \times
\sum_{j=1}^3 U_{ej} U_{ej}^* \int_{-\infty}^{\infty} \dd \delta Q 
\exp \left[ 
-\frac{i\delta Q}{\hbar} \left( T - \Delta t_j \right) \right]
\times
\frac{\sin\left( \frac{\tau}{2\hbar}(1-\beta)\delta Q \right)}%
{\pi (1-\beta)\delta Q}
\label{factor} \\ && \times 
\exp \left[ -i \left( 
\frac{\bar Q}{\hbar} \left(t_D - t_S - \frac{L}{c} \right) +
\frac{(m_j c^2)^2 L}{2 \hbar c \bar Q}
\right) \right],
\label{A-approx1}
\end{eqnarray}
\end{subequations}
where we have defined 
\begin{equation}\label{tdef}
T = t_M - t_S - \frac{L}{c} + \beta \left( t_D - t_M \right)
\quad \mbox{and} \quad
\Delta t_j = \frac{(m_j c^2)^2 L}{2c {\bar Q}^2}.
\end{equation}

\paragraph{Integration over $\delta Q$:} 
The integral in equation~(\ref{A-int}) can now be performed exactly 
by using theorem~\ref{int} of the appendix. The result is
\begin{subequations}\label{Aab}
\begin{eqnarray}
\mathcal{A}_\infty &\simeq& 
\frac{1}{L(1-\beta)}\, \bar{\mathcal{A}}_S \bar{\mathcal{A}}_D
\sum_{j=1}^3 U_{ej} U_{ej}^*\: 
\Theta_\tau\left( \frac{T - \Delta t_j}{1-\beta} \right)
\label{a} \\ && \times 
\exp \left[ -i \left( 
\frac{\bar Q}{\hbar} \left(t_D - t_S - \frac{L}{c} \right) +
\frac{(m_j c^2)^2 L}{2 \hbar c \bar Q}
\right) \right].
\label{b}
\end{eqnarray}
\end{subequations}

Now we turn to a discussion of the quality of our approximations 
in the integration over $\delta Q$. Let us reconsider in equation~(\ref{A-int}) 
finite boundaries $\pm \Delta Q/2$ of the integral. 
We know that the width $\sigma_D$ of $\psi_D$ provides a kind of 
cut-off for the $Q$-integration. Moreover, for $Q \sim \sigma_D$ 
the specific form of $\psi_D$, which we do not really know, would come into 
play. Therefore, a $\Delta Q$ fulfilling 
$2\pi\hbar/\tau \ll \Delta Q \ll \sigma_D$ 
would give an excellent approximation 
of the integral without the need to know $\psi_D$. With $\sigma_D \sim 10$\,keV 
and $2\pi\hbar/\tau \sim 4 \times 10^{-9}$\,eV for $\tau = 10^{-6}$\,s 
there is ample space for such a $\Delta Q$.
However, in order to be able 
to compute the integral over $\dd\delta Q$ explicitly, 
we have taken the limit $\Delta Q \to \infty$. Therefore, we have to 
check which error we make by taking this limit. 
As discussed in the appendix, the integration over $\delta Q$ amounts to 
two integrals of the form 
\begin{equation}
I \equiv \int_{-\Delta Q/2}^{\Delta Q/2} \dd \delta Q \,
\frac{\sin\frac{t\delta Q}{2\hbar}}{2\pi \delta Q}
\quad \mbox{with} \quad 
t = \left| \frac{T - \Delta t_j}{1-\beta} \pm \tau/2 \right|.
\end{equation}
Denoting by $1/f$ the error we make in $I$ 
by extending the boundaries to infinity
and using equation~(\ref{int5}), we find
\begin{equation}\label{error}
\frac{1}{f} \gtrsim \frac{8\hbar}{\pi t\Delta Q}
\quad \Rightarrow \quad
\Delta Q \gtrsim 1.7 \times 10^{-9}\,\mbox{eV} \times f \times 
\frac{10^{-6}\,\mathrm{s}}{t}.
\end{equation}
We see that it is possible to have $\Delta Q \ll \sigma_D$ even for 
very small errors like $1/f = 10^{-6}$, provided $t$ is not too small. 
However, for $t \to 0$, the boundary $\Delta Q$
approaches $\sigma_D$ and, for very small regions around the 
jump discontinuities of $\Theta_\tau$, 
equation~(\ref{a}) might not be a good approximation.

\paragraph{Physical interpretation of the result for $\mathcal{A}_\infty$:}
Clearly, the phases in equation~(\ref{b}) lead to neutrino oscillations.

So we turn to the physical interpretation of $\Theta_\tau$
in equation~(\ref{a}). We note that the time interval
$[t_M - \tau/2,\, t_M + \tau/2]$ when the detector takes data is 
\emph{specified} and, therefore, equation~(\ref{a}) makes a statement about 
the time $t_S$ when the neutrino is \emph{produced}. That statement 
is to be dervied from
\begin{equation}
-\frac{\tau}{2} < \frac{T - \Delta t_j}{1-\beta} < \frac{\tau}{2}.
\end{equation}
Actually, this equation leads to three intervals, 
one for each massive neutrino:
\begin{equation}\label{t_S}
t_M + \beta \left( t_D - t_M \right) - \frac{L}{c} - \Delta t_j 
- \frac{\tau}{2} (1-\beta) 
< t_S < 
t_M + \beta \left( t_D - t_M \right) - \frac{L}{c} - \Delta t_j 
+ \frac{\tau}{2} (1-\beta).
\end{equation}
Now it is interesting to ask the question whether this abstract QFT result 
has a simple physical interpretation.

Firstly we consider the quantity $\Delta t_j$. This is the only term in 
equation~(\ref{t_S}) which contains a neutrino mass $m_j$. It is easy to see 
that an ultrarelativistic particle with mass $m_j$ covers 
the distance $L$ in the time $L/c + \Delta t_j$. 
Therefore, $\Delta t_j$ is the time delay due to the neutrino velocity 
being slightly smaller than the velocity of light. 
In other words, given the time when it is measured, it must be produced earlier 
than a massless particle by the time $\Delta t_j$.\footnote{A similar 
conclusion has been reached in~\cite{kobach}.}

Secondly we demonstrate that in equation~(\ref{t_S}) the remaining terms have
a nice classical interpretation as well.
It is sufficient to consider a one-dimensional setting.
We know that the detector particle is at the location $x_D$ 
at time $t_D$ and moves with velocity 
$v = \beta c$ defined in equation~(\ref{beta}). 
In addition, having already taken into account $\Delta t_j$, 
the neutrino propagates with velocity $c$ and 
is located at $x_S$ where it is produced at time $t_S$.
Therefore, we have the trajectories 
\begin{equation}
x_D(t) = (t-t_D) v + x_D 
\quad \mbox{and} \quad
x_\nu(t) = (t-t_S) c + x_S
\end{equation}
for the detector particle and the neutrino, respectively.
We furthermore stipulate that the neutrino is measured at time $t'_M$.
Classically, $t'_M$ is determined by the equation
\begin{equation}
x_\nu(t'_M) = x_D(t'_M) \quad \Rightarrow \quad 
t'_M = \frac{1}{1-\beta} \left( t_S - \beta t_D + \frac{L}{c} \right), 
\end{equation}
where we have set $L = x_D - x_S$.
Since $t'_M$ must be inside the measurement interval, we obtain the inequalities
\begin{equation}
t_M - \frac{\tau}{2} < 
\frac{1}{1-\beta} \left( t_S - \beta t_D + \frac{L}{c} \right) < 
t_M + \frac{\tau}{2},
\end{equation}
which are equivalent to those of equation~(\ref{t_S}) without $\Delta t_j$. 

In a nutshell, we have thus found that the $\Theta_\tau$-term in 
equation~(\ref{Aab}), albeit derived from QFT, can be nicely interpreted in the 
classical sense. This strengthens our 
confidence in the QFT model presented here.

Note that in principle the occurrence of $\Delta t_j$ in equation~(\ref{a}) 
is a source of decoherence in the \emph{amplitude}, if 
$\tau \lesssim \Delta t_j$.\footnote{In the wave-packet picture 
this corresponds to the decoherence due to the separation of 
the wave packets pertaining to neutrinos with different masses moving therefore
with slightly different velocities.} 
However, in practice this source of decoherence 
is irrelevant. To give a numerical example, we set $\bar Q = 0.5$\,MeV, 
$m_j = 0.1$\,eV and $L = 300$\,km. 
Then $\Delta t_j \simeq 2 \times 10^{-17}$\,s and for all 
practical purposes we have $\tau \gg \Delta t_j$.

\paragraph{Averaging over $t_D$:}
\begin{figure}[t]
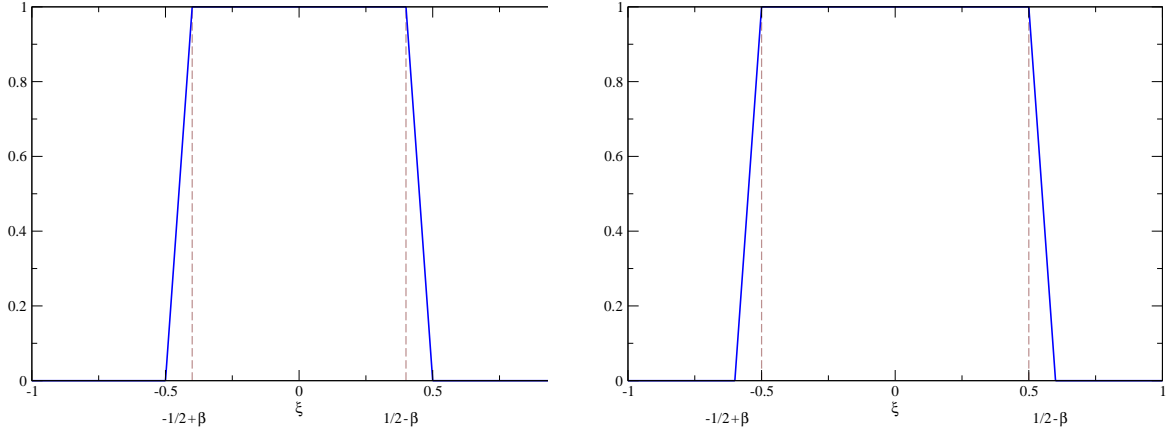

\begin{center}
\begin{tabular}{cc}
\epsfig{file=figplus.eps,width=7.5cm}
&
\epsfig{file=figminus.eps,width=7.5cm}
\end{tabular}
\end{center}
\caption{The function $F_\beta(T'/\tau)$ plotted against  
$\xi = T'/\tau$ for $\beta = 0.1$ (left plot) and 
$\beta = -0.1$ (right plot).
Note that this value of $\beta$ is grossly 
exaggerated for the purpose of giving a visible effect in the figure.}
\label{figure}
\end{figure}
As already stated in section~\ref{example}, it is necessary to average 
in the cross section over the time $t_D$---see equation~(\ref{equivalent}).
In the following we neglect $\Delta t_j$.
Since $\left( \Theta_\tau \right)^2 = \Theta_\tau$, this amounts to 
\begin{equation}
\frac{1}{\tau} \int_{t_M-\tau/2}^{t_M+\tau/2} \dd t_D\, 
\Theta_\tau \left( \frac{T}{1-\beta} \right) \equiv F_\beta(T'/\tau) 
\quad \mbox{with} \quad 
T' = t_M - t_S - L/c.
\end{equation}
Because of the rectangular form of $\Theta_\tau$ we obtain the 
trapezoid form of $F_\beta(T'/\tau)$ displayed in 
figure~\ref{figure}.\footnote{Here we have assumed that the tiny intervals 
of $T$ around $\pm (1 - \beta)\tau/2$, 
where $\Theta_\tau\left(\frac{T}{1-\beta}\right)$ is possibly not a good 
description of the time-dependence of the amplitude, 
does not affect the averaging.}
Since $|\beta| \ll 1$ this averaging has no practical effect.

\paragraph{Final approximation for the amplitude:}
Now we take into account that for all practical purposes we can neglect both
$\Delta t_j$ and $\beta$, as we have demonstrated above. 
Removing in addition the unphysical part of the phase in equation~(\ref{b})
and denoting the thus obtained approximate amplitude by $\mathcal{A}'_\infty$, 
we arrive at the exceedingly simple formula
\begin{equation}\label{main}
\mathcal{A}'_\infty = 
\frac{1}{L}\, \bar{\mathcal{A}}_S \bar{\mathcal{A}}_D
\sum_{j=1}^3 U_{ej} U_{ej}^*\: \Theta_\tau\left( t_M - t_S - \frac{L}{c} \right)
\exp \left[ -i \frac{(m_j c^2)^2 L}{2 \hbar c \bar Q} \right].
\end{equation}
This is the main result of our paper.

\section{Summary and conclusions}
\label{conclusions}
\paragraph{Summary of assumptions:}
\begin{enumerate}
\item\label{compound}
The process of neutrino production and detection is considered as 
one compound process, with the neutrino source located around $\vec x_S$ and 
detection around $\vec x_D$ such that the two locations are separated by a
macroscopic distance $L = |\vec x_D - \vec x_S|$.
\item\label{cross section}
Neutrino oscillation probabilities are to be derived from the 
cross section of the compound process in the standard way, \textit{i.e.}\ with 
all final particles in production and detection represented by plane waves.
\item\label{thermal}
The detector particle is at rest apart from thermal motion; 
it is assumed that the latter gives 
the correct order of magnitude of its energy spread.\footnote{Detector 
particles are bound in atoms and molecules and neighbouring bound states 
will influence the thermal behaviour of the detector particle.}
\item\label{nu detection}
For detecting a neutrino event and separating it from the background 
an energy of at least several 100\,keV has to be deposited in the detector.
\end{enumerate}
Concerning items~\ref{compound} and~\ref{cross section}, 
we have studied the amplitude leading to the 
cross section in the asymptotic limit $L \to \infty$. To do so we have 
taken advantage of 
theorem~\ref{limL} proven in~\cite{GS}.
As for item~\ref{cross section}, 
this is our essential assumption and agrees with the 
computation of cross sections in particle-physics textbooks. We think that 
item~\ref{thermal} contains a reasonable assumption for realistic experiments, 
and the same applies to item~\ref{nu detection}.  

\paragraph{What is new in the present paper:}
In~\cite{G2003,GS,GSM} it has been assumed that the detector particle is 
in an energy eigenstate. In the present paper we have relaxed this assumption 
and we have allowed for a finite time interval $\tau$ 
for the measument in the neutrino detector, as practised in 
oscillation experiments with neutrino sources generated by accelerators. 

\paragraph{The limit $\tau \to \infty$:}
Nevertheless in this limit we obtain 
for all practical purposes the same result as in~\cite{G2003,GS,GSM}:
\begin{enumerate}
\renewcommand{\labelenumi}{\roman{enumi})}
\item
For each neutrino mass $m_j$, 
the neutrino energy $Q_j$, equation~(\ref{Q_j}), 
is completely fixed by the detection process.
However, the dependence of $Q_j$ on $m_j$, which leads to a tiny amount of 
neutrino oscillations in time in addition to those in space, is totally 
negligible and for all practical purposes we have a common neutrino 
energy $\bar Q$, equation~(\ref{Qbar1}), leading to oscillations in space only.
\item
There is no decoherence effect in the amplitude $\mathcal{A}_\infty$.
Therefore, the only source of decoherence is the incomplete 
knowledge of the final momenta in the 
detection cross section.\footnote{Under simplifying assumptions this has 
also been found in~\cite{kobach}.}
\end{enumerate}
Let us repeat the logic leading to statements~i) and~ii). 
In the limit $\tau \to \infty$ 
we obtain a $\delta$-function in energy for the 
detection process---see equation~(\ref{tau-infty})---and 
in the asymptotic limit 
$L \to \infty$ the direction of the neutrino momentum is fixed by 
$\vec\ell$ of equation~(\ref{vec ell}),
which leaves, neglecting $m_j$, only $\bar Q$ to be determined by 
equation~(\ref{Qbar1}) in terms 
of \emph{final momenta}. 
The summation over the final 
momenta occurs in the cross section---in the standard way of computing 
it in particle physics; therefore, final momenta (and energies) have no bearing
on the coherence issue in the amplitude we are concerned 
with in our paper.

\paragraph{Finite $\tau$:}
If $\tau$ is not infinitely large, energy conservation in the detection process 
is not exact but this has no bearing on the utmost negligibility of $m_j$ in 
the kinematics of this process. Now the neutrino energy $Q$ can vary within 
an interval of approximate length $\Delta Q \sim 2\pi \hbar/\tau$ around 
a mean value $\bar Q$ determined by equation~(\ref{Qbar1}). 
Note that $\Delta Q$ is an expression of the time--energy uncertainty principle 
due to the finite measurement time interval $\tau$.
According to the QFT formalism we still have to perform the integration over 
$Q$ in the interval $[\bar Q - \Delta Q/2,\,\bar Q + \Delta Q/2]$, the last 
integration in computing the compound amplitude of neutrino 
production and detection. And exactly this integration leads to the 
time correlation between neutrino source and neutrino detection, 
expressed by the function $\Theta_\tau$ in $\mathcal{A}_\infty$, 
equation~(\ref{Aab}), that one would expect on physical grounds.

Furthermore, neutrino oscillations still happen with the same energy 
$\bar Q$ as before when the limit $\tau \to \infty$ was considered, without any 
decoherence for realistic $\tau$---see, however, the discussion 
on $\tau$ and $\Delta t_j$ in the previous section. 
We emphasize that the QFT formalism 
naturally leads to oscillations in space but not in time, which 
confirms the qualitative analysis 
in~\cite{lipkin1,stodolsky,lipkin2}.\footnote{Interpreting 
$L/c$ as the time of flight 
and defining momenta $P_j = \bar Q/c - (m_j c^2)^2/(2c\bar Q)$, one could 
rewrite the phases in equation~(\ref{b}) such that oscillations occur in time 
with different momenta, but this would be an unnecessary and artificial 
reformulation without physical content, because the true time information 
is contained in $\Theta_\tau$.}

Finally, we emphasize that our analysis corroborates 
the importance of the neutrino detection process for questions concerning 
coherence in the neutrino oscillation amplitude~\cite{weiss,kiers}.

\paragraph{Conclusions:}
In summary, we have demonstrated that QFT
naturally leads to the standard neutrino oscillations in vacuum \emph{and} 
the time correlation between neutrino production and detection, provided 
we take into account experimental conditions and compute the 
cross section of the compound neutrino production--detection process in 
the standard way, \textit{i.e.}\ using planes waves for the final states.
Although for definiteness we have assumed specific reactions for neutrino 
production and detection, our conclusions are general. 
In effect, we have found that 
neutrino oscillations take place in space with a single energy 
$\bar Q$. This is in contradiction to the 
wave-packet picture,\footnote{In the wave packet picture 
oscillations take place in both space and time.}
but we think in our framework this conclusion is compelling.
Of course this contradiction is only a matter of theoretical interest and 
not of practical 
relevance because in good approximation both frameworks give the same 
oscillation probabilities. In particular, the errors in the 
determination of the neutrino energies lead to decoherence effects in 
the oscillation \emph{probabilities} which override all subtleties in 
the oscillation \emph{amplitudes} discussed in the present paper.

\vspace{3mm}

\paragraph{Acknowledgements:}
The author thanks H.\ Neufeld for clarifying discussions in the 
early stage of the paper and he is very grateful to S.M.\ Bilenky and 
A.\ Olshevskiy for information on the time synchronization between 
the proton spill time and neutrino detection 
in accelerator neutrino-oscillation experiments.

\newpage

\appendix

\setcounter{equation}{0}
\renewcommand{\theequation}{A.\arabic{equation}}

\section{Appendix}

Firstly, we recapitulate the theorem proven in~\cite{GS}, 
which we need in the context of the integration over $\dd^3 q$.
\begin{theorem}\label{limL}
Let $\Phi: \mathbbm{R}^3 \to \mathbbm{R}^3$ be a three times continuously 
differentiable function of the variable $\vec q$ such that $\Phi$ itself and 
all its first and second derivatives decrease at least like 
$1/{\vec q}^{\,2}$ for $|\vec q\,| \to \infty$, $A$ a real number and 
\begin{equation}
J(\vec L\,) \equiv \int \dd^3 q\, \Phi(\vec q\,)\, e^{-i\vec q \cdot \vec L}\,
\frac{1}{A - {\vec q}^{\,2} + i\epsilon}.
\end{equation}
Then in the asymptotic limit $L \equiv |\vec L\,| \to \infty$ one obtains,
for $A > 0$,
\begin{equation}
J(\vec L\,) = -\frac{2\pi^2}{L}\, \Phi \left( -\sqrt{A}\, \vec L/L \right)
\, e^{i \sqrt{A} L} + \mathcal{O} \left( L^{-3/2} \right),
\end{equation}
whereas for $A < 0$ the integral decreases like $L^{-2}$.
\end{theorem}

Secondly, we compute a simple integral, used in the body of the paper
for the integration over $\dd q^0$.
\begin{theorem}\label{int}
Let $\tau$ be positive and $T \in \mathbbm{R}$. Then
\begin{equation}\label{int2}
\int_{-\infty}^\infty \dd u\, \exp\left( iTu \right)\, 
\frac{\sin \left( \frac{1}{2} \tau u \right)}{\pi u} = 
\left\{ \begin{array}{ccc} 
1 & \mbox{for} & |T| < \frac{1}{2} \tau, 
\\[2mm]
0 & \mbox{for} & |T| > \frac{1}{2} \tau.
\end{array}
\right.
\end{equation}
\end{theorem}
\noindent \textbf{Proof:}
We take the function $\Theta_\tau(t)$ defined in equation~(\ref{Dt}) and 
compute its Fourier transform 
\begin{equation}
\widetilde \Theta_\tau(u) = \frac{1}{\sqrt{2\pi}}\,
\int_{-\infty}^\infty \dd t \, e^{-iut}\, \Theta_\tau(t) = 
\sqrt{\frac{2}{\pi}}\, \frac{\sin\left( \frac{1}{2} \tau u \right)}{u}.
\end{equation}
Application of the inverse Fourier transform then leads to 
\begin{equation}
\Theta_\tau(T) = \frac{1}{\sqrt{2\pi}}\,
\int_{-\infty}^\infty \dd u \, e^{iTu} \widetilde\Theta_\tau(u) = 
\int_{-\infty}^\infty \dd u \, e^{iTu} \, 
\frac{\sin \left( \frac{1}{2} \tau u \right)}{\pi u},
\end{equation}
which is the desired result.

Actually, equation~(\ref{int2}) is not only true in the sense of the 
Fourier transform on $L^2(\mathbbm{R})$, it also holds pointwise. 
No special precautions have to be taken for the convergence of the 
improper integral because both integrals 
\begin{equation}\label{int3}
\lim_{a \to \infty} \int_{-a}^0 \dd u\, \exp\left( iTu \right)\, 
\frac{\sin \left( \frac{1}{2} \tau u \right)}{\pi u}
\quad \mbox{and} \quad
\lim_{b \to \infty} \int_0^b \dd u\, \exp\left( iTu \right)\, 
\frac{\sin \left( \frac{1}{2} \tau u \right)}{\pi u}
\end{equation}
exist for $|T| \neq \tau/2$. However, 
if $|T| = \tau/2$, then both integrals in equation~(\ref{int3}) diverge.
A simple remedy, which also makes sense for $|T| = \tau/2$, is given by 
\begin{eqnarray}
\lim_{a \to \infty} \int_{-a}^a \dd u\, \exp\left( iTu \right)\, 
\frac{\sin \left( \frac{1}{2} \tau u \right)}{\pi u} &=&
\nonumber \\ 
\lim_{a \to \infty} \int_{-a}^a \frac{\dd u}{2\pi u} \left[
\sin\left( (T + \tau/2)u \right) - 
\sin\left( (T - \tau/2)u \right) \right] &=& 
\left\{ \begin{array}{ccc} 
1 & \mbox{for} & |T| < \frac{1}{2} \tau, 
\\[2mm]
\frac{1}{2} & \mbox{for} & |T| = \frac{1}{2} \tau, 
\\[2mm]
0 & \mbox{for} & |T| > \frac{1}{2} \tau.
\end{array}
\right.
\label{int4}
\end{eqnarray}

Let us now estimate the rate of convergence of the two integrals in 
equation~(\ref{int4}). First we observe that for all 
$r \in \mathbbm{N}$ the inequalities 
\begin{equation}
-\frac{2}{\pi} \left( \frac{1}{2r - 1} - \frac{1}{2r + 1} \right) < 
\int_{(2r - 1)\pi}^{(2r + 1)\pi} \dd y \,\frac{\sin y}{y} < 0
\end{equation}
hold due to $1/y$ being monotonously decreasing and 
$\int_0^\pi \dd y \sin y = 2$. Then, for 
\begin{equation}
t = |T \pm \tau/2|
\end{equation}
and 
setting 
\begin{equation}
ta = (2r - 1)\pi 
\quad  \mbox{and} \quad
y = tu
\end{equation}
we obtain
\begin{equation}\label{int5}
-\frac{2}{\pi ta} < 
\left( \int_{-\infty}^{-a} + \int_{a}^\infty \right) \dd u \,
\frac{\sin tu}{2\pi u} =
\left( \int_{-\infty}^{-ta} + \int_{ta}^\infty \right) \dd y \,
\frac{\sin y}{2\pi y} < 0.
\end{equation}
We see that with $t \to 0$ the convergence of the integral worsens. This 
reflects the discontinuity in $t$:
\begin{equation}
\int_{-\infty}^\infty \dd u \, \frac{\sin tu}{2\pi u} = 
\left\{ \begin{array}{ccc}
1/2 & \mbox{for} & t > 0, \\[2mm]
0 & \mbox{for} & t = 0.
\end{array} \right.
\end{equation}

\end{document}